\documentclass[12pt]{article}
 \usepackage{epsf,latexsym}
\begin{document}

\title{Charged SU(N) Einstein-Yang-Mills Black Holes}
\vspace{1.5truecm}
\author{
{\bf Burkhard Kleihaus, Jutta Kunz and Abha Sood}\\
Fachbereich Physik, Universit\"at Oldenburg, Postfach 2503\\
D-26111 Oldenburg, Germany}

\date{\today}

\maketitle
\vspace{1.0truecm}

\begin{abstract}
We consider asymptotically flat static
spherically symmetric black hole solutions in $SU(N)$ 
Einstein-Yang-Mills theory.
Embedding the $N$-dimensional representation of $su(2)$ in $su(N)$,
the purely magnetic gauge field ansatz contains $N-1$ functions.
When one or more of these gauge field functions
are identically zero, magnetically charged EYM black hole solutions emerge,
consisting of a neutral and a charged gauge field part, 
based on non-abelian subalgebras and the Cartan subalgebra of $su(N)$,
respectively.
We classify these charged black hole solutions in general
and present numerical solutions for $SU(5)$ EYM theory.

\end{abstract}

\vfill
\noindent {Preprint hep-th/9705179} \hfill\break
\vfill\eject
\section{Introduction}

$SU(2)$ Einstein-Yang-Mills (EYM) theory possesses non-abelian
static spherically symmetric
globally regular and black hole solutions
\cite{bm,bh}.
These solutions are unstable \cite{volkov4} and carry no charge. 
With non-abelian hair outside their regular event horizon,
the black hole solutions represent counterexamples 
to the ``no-hair conjecture''.

The $SU(2)$ solutions, based on a purely magnetic ansatz, are labelled
by the node number $n$ of the gauge field function $u$.
For fixed horizon radius $x_{\rm H}$
and increasing node number $n$ the sequence 
of neutral $SU(2)$ EYM black hole solutions tends to a limiting
magnetically charged solution.
When $x_{\rm H} \ge 1$,
this limiting solution is an embedded 
Reissner-Nordstr\o m (RN) solution \cite{yasskin}.
It has a vanishing gauge field function $u$
and the RN metric has
magnetic charge $P=1$ 
\cite{lim,kks3}.

The non-abelian static spherically symmetric
globally regular and black hole solutions
of $SU(3)$ Einstein-Yang-Mills (EYM) theory are obtained by
embedding the $3$-dimensional representation of $su(2)$ in $su(3)$.
The purely magnetic gauge field ansatz then involves
two gauge field functions, $u_1$ and $u_2$,
and the solutions are labelled
by the corresponding node numbers $(n_1,n_2)$
\cite{kuenzle,kks2,kks3}.

The neutral $SU(3)$ EYM solutions form sequences $(i,i+n)$, with $i$ fixed.
In the limit $n \rightarrow \infty$,
these sequences of neutral $SU(3)$ EYM solutions tend to
limiting solutions, carrying magnetic charge of norm
$P=\sqrt{3}$ \cite{kks2,kks3}.
When $x_{\rm H} \ge P$,
the limiting solutions are magnetically charged non-abelian 
black hole solutions \cite{gv}, in which
one of the two gauge field functions is identically zero.
The gauge field of these limiting solutions consists of two parts,
a non-abelian $su(2)$ part and a $su(3)$ Cartan subalgebra (CSA) part.
The field strength tensor component $F_{\theta\phi}$ of the
CSA part of the gauge field
does not vanish identically asymptotically,
yielding a CSA magnetic charge.
In contrast, the field strength tensor of the $su(2)$ part
of the gauge field decays to zero asymptotically,
yielding no magnetic charge.
When both gauge field functions are identically zero,
embedded RN solutions are obtained,
carrying magnetic CSA charge of norm $P=\sqrt{4}$ \cite{yasskin}.

In $SU(2)$ EYM theory all static spherically symmetric
EYM black hole solutions with non-zero charge are embedded
RN solutions \cite{ersh,popp}.
In contrast, this ``non-abelian baldness theorem'' no longer holds
for $SU(3)$ EYM theory, 
which does allow for black hole solutions,
with magnetic CSA charge 
and with non-abelian $su(2)$ gauge field configurations \cite{gv}.
$SU(3)$ EYM theory additionally allows for black hole solutions with 
both electric and magnetic CSA charge
and with non-abelian $su(2)$ gauge field configurations \cite{gv},
based on a more general ansatz not considered here.

Here we consider static spherically symmetric $SU(N)$ EYM solutions, based on
the purely magnetic gauge field ansatz, obtained by
embedding the $N$-dimensio\-nal representation of $su(2)$ in $su(N)$.
The ansatz contains $N-1$ gauge field functions.
When one or more of these functions
are identically zero, magnetically charged EYM black hole solutions emerge.
Their gauge fields
consist in general of non-abelian $su(\bar{N})$ ($\subset su(N)$) parts
and a $su(N)$ CSA part.
Only when all gauge field functions are identically zero,
embedded RN solutions emerge.
We here present a general analysis of the possible types 
of these magnetically charged $SU(N)$ EYM black hole solutions.
We give the full classification
for the example of $SU(5)$ EYM theory,
and construct several sequences 
of magnetically charged non-abelian black hole solutions numerically.

\section{SU(N) EYM Equations of Motion}

We consider the $SU(N)$ Einstein-Yang-Mills action
\begin{equation}
S=S_G+S_M=\int L_G \sqrt{-g} d^4x + \int L_M \sqrt{-g} d^4x
\ \label{action}  \end{equation}
with
\begin{equation}
L_G=\frac{1}{16\pi G}R \ , \ \ \
L_M=-\frac{1}{2} {\rm Tr} (F_{\mu\nu} F^{\mu\nu})
\ , \label{lagm} \end{equation}
and with field strength tensor
$F_{\mu\nu}= \partial_\mu A_\nu - \partial_\nu A_\mu
            - i e [A_\mu,A_\nu]$,
gauge field
$A_\mu = \frac{1}{2} \lambda^a A_\mu^a$
and gauge coupling constant $e$.
Variation of the action eq.~(\ref{action}) with respect to the metric
$g^{\mu\nu}$ leads to the Einstein equations,
and variation with respect to the gauge field $A_\mu$ 
leads to the matter field equations.

To construct static spherically symmetric black hole solutions
we employ Schwarz\-schild-like coordinates and adopt
the spherically symmetric metric
\begin{equation}
ds^2=g_{\mu\nu}dx^\mu dx^\nu=
  -{\cal A}^2{\cal N} dt^2 + {\cal N}^{-1} dr^2 
  + r^2 (d\theta^2 + \sin^2\theta d\phi^2)
\ , \label{metric} \end{equation}
with the metric functions ${\cal A}(r)$ and 
${\cal N}(r)=1-\frac{2m(r)}{r}$.

The static spherically symmetric ans\"atze
for the gauge field $A_{\mu}$ of $SU(N)$ EYM theory
are based on the $su(2)$ subalgebras of $su(N)$.
Here we do not consider all inequivalent embeddings of $su(2)$ in $su(N)$
but restrict ourselves to the embedding
of the $N$-dimensional representation of $su(2)$.
The corresponding ansatz is given by \cite{kuenzle}
\begin{equation} \label{ansa}
 A_{\mu}^{(N)}  dx^\mu   =  \frac{1}{2e} \left( 
\begin{array}{ccccc}
(N-1)\cos\theta d\phi & \omega_1 \Theta & 0 & \ldots & 0 \\
\omega_1 \bar \Theta & (N-3)\cos\theta d\phi & \omega_2 \Theta & 
\ldots & 0 \\
\vdots & & \ddots & & \vdots \\
0 & \ldots & 0 & \omega_{N-1} \bar \Theta & (1-N)\cos\theta d\phi
\end{array} \right)
\label{amu} \end{equation}   
with $\Theta = i d \theta + \sin \theta d \phi$,
and $A_0=A_r=0$.
The ansatz contains $N-1$ matter field functions
$\omega_j(r)$, $j=1,\ldots,N-1$,
and leads to the field strength tensor components
$F_{r\theta}=\partial_r A_\theta$,
$F_{r\phi}=\partial_r A_\phi$ and
\begin{equation} \label{F1}
F_{\theta\phi}=(1/2e) {\rm diag} (f_1,...,f_N) \sin \theta
\   \end{equation}   
with
\begin{equation} 
f_j = \omega_j^2 - \omega^2_{j-1} + \delta_j \ , \ \ \
\delta_j = 2j - N -1 \ , \ \ \ j=1,\ldots,N 
\  \ \ \ (\omega_0=\omega_N=0)
\ . \label{f1} \end{equation}   

Let us now introduce the dimensionless coordinate 
$x=er/{\sqrt{4\pi G}}$,
the dimensionless mass function
\begin{equation}
\mu=\frac{em}{\sqrt{4\pi G}} 
\  , \end{equation}
and the scaled matter field functions \cite{kuenzle}
\begin{equation}
 u_j = \frac{\omega_j}{\sqrt{\gamma_j}} \ , \ \ \
 \gamma_j = {j (N - j) } 
\  . \end{equation}
The Einstein equations then yield for the metric functions
the equations
\begin{eqnarray}
\mu^{'} & = & {\cal N}\cal{G}  + \cal{P}  \quad , \label{sunm}\\
\frac{{\cal A}^{'}}{{\cal A}} & = & \frac{2 {\cal G}}{x} 
\ , \label{suna} \end{eqnarray}
(the prime indicates the derivative with respect to $x$),
where
\begin{equation}
{\cal G} = \sum_{j=1}^{N-1} \gamma_j u_j^{' 2} \ , \ \ \
{\cal P} = \frac{1}{4 x^2} \sum_{j=1}^N f_j^2
\   \end{equation}
and $f_j = \gamma_j u_j^2 - \gamma_{j-1} u^2_{j-1} + \delta_j$
(see eq.~(\ref{f1})).
The equations for the matter field functions are
\begin{equation}
({\cal A}{\cal N}u_j^{'})^{'} 
 + \frac{1}{2x^2} {\cal A} (f_{j+1} - f_j) u_j = 0
\ , \label{sunu} \end{equation}
where the metric function ${\cal A}$ can be eliminated by means of
eq.~(\ref{suna}).
We note the symmetry of the equations with respect to
the transformation
$u_j \rightarrow u_{N-j}$, $j=1,\ldots,N-1$.

\section{SU(N) EYM Solutions}

Our aim here is to classify the
charged $SU(N)$ EYM black hole solutions,
which appear as limiting solutions of sequences of neutral
non-abelian $SU(N)$ solutions, based on the $N$-dimensional embedding
of $su(2)$ in $su(N)$, eq.~(\ref{amu}).

\subsection{Neutral SU(N) EYM Solutions}

We begin by briefly considering the neutral $SU(N)$ EYM solutions
and their boundary conditions. In these globally regular
or black hole solutions all $N-1$ gauge field functions are non-trivial.
(In the trivial case, where all gauge field functions are identical to one,
Schwarzschild solutions are obtained.)
For asymptotically flat solutions the metric functions ${\cal A}$
and $\mu$ must approach constants at infinity.
The time coordinate is fixed by ${\cal A}(\infty)=1 $, 
and the mass of the solutions is given by $\mu(\infty)$.
The gauge field functions $u_j$ satisfy \cite{kuenzle}
\begin{equation}
u_j(\infty)=\pm 1 \ , \ \ \ j=1,\ldots,N-1
\ . \label{bc1} \end{equation}
The field strength tensor component  
$F_{\theta \phi}^{({N})}$ decays like $O(r^{-1})$ asymptotically,
whereas the components $F_{r\theta}^{({N})}$ and $F_{r\phi}^{({N})}$
decay like $O(r^{-2})$.
Therefore these solutions are magnetically neutral.
For globally regular solutions the boundary conditions at the origin 
are $\mu(0)=0$ and \cite{kuenzle}
\begin{equation}
u_j(0)=\pm 1 \ , \ \ \ j=1,\ldots,N-1
\ . \label{bc3} \end{equation}
For black hole solutions with a regular horizon 
with radius $x_{\rm H}$, the boundary conditions at the horizon
are 
${\cal N}(x_{\rm H})=0$, i.~e.~$\mu(x_{\rm H})= x_{\rm H}/2$,
and
\begin{equation}
 \left. {\cal N}^{'}u_j^{'} + \frac{1}{2x^2}   (f_{j+1} - f_j) u_j 
\right|_{x_{\rm H}} =0
\ . \label{bc5} \end{equation}

For $SU(2)$ these neutral globally regular and black hole 
solutions are discussed in \cite{bm,bh},
for $SU(3)$ in \cite{kuenzle,kks2,kks3}
and for $SU(4)$ in \cite{kksw}.
The solutions are labelled by the node numbers
$n_j$ of the functions $u_j$.
When the node numbers of one or more gauge field functions tend to infinity,
the solutions approach limiting solutions, carrying magnetic CSA charge 
of norm $P$.
Considering black hole solutions with $x_{\rm H} > P$
(or the exterior part of the solutions with $x_{\rm H} < P$)\footnote{
For globally regular solutions and black hole solutions
with $x_{\rm H} < P$ the limiting solutions consist of two
parts: an interior part for $x<P$ and an exterior part for $x>P$
\cite{lim,kks3,kksw}.},
we observe that 
in these limiting solutions the corresponding gauge field functions
are identically zero.

\subsection{Charged SU(N) EYM Solutions}

When one or more of the $N-1$ gauge field functions 
are identically zero,
magnetically charged solutions are obtained.
To classify the charged black hole solutions
obtained within the ansatz (\ref{amu}),
let us first assume that precisely one gauge field function 
is identically zero, $\omega_{j_{1}} \equiv 0$.
The ansatz for the gauge field then splits into two parts
\begin{equation}
\begin{array}{ccc}
A_\mu^{(N)} dx^\mu & = & 
\left( 
\begin{array}{cc}
{\rm \fbox{$ A_\mu^{(j_1)} dx^\mu  $}} &                      \\
                     &  {\rm \fbox{$A_\mu^{(N-j_1)} dx^\mu $}}\\
\end{array}
\right)
 + {\cal H}_{j_1}
\end{array} \vspace{1.cm} 
\ \label{amu1} \end{equation}
with  
${\cal H}_{j_1}  =  \frac{\cos\theta d\phi}{2 e} h_{j_1}$ and
\begin{equation}
\begin{array}{ccc}
{h}_{j_1} & = & 
\left( 
\begin{array}{cc}
{\rm \fbox{$(N-j_1){\bf 1}_{(j_1)} $}}&     \\
      &{\rm \fbox{$ -j_1 {\bf 1}_{(N-j_1)}$}}   \\            
\end{array}
\right)\\
\end{array}\vspace{1.cm} 
\ . \label{h1} \end{equation}
$A_\mu^{(j_1)}$ and $A_\mu^{(N-j_1)}$ denote the non-abelian
spherically symmetric ans\"atze for the $su(j_1)$
and $su(N-j_1)$ subalgebras of $su(N)$
(based on the $j_1$ and $(N-j_1)$-dimensional embeddings of $su(2)$,
respectively), referred to by $su(\bar{N})$ in the following.
${\cal H}_{j_1}$ represents the ansatz for the element 
${h}_{j_1}$ of the CSA of $su(N)$.
The field strength tensor splits accordingly into 
\begin{equation}
\begin{array}{ccc}
F_{\mu\nu}^{(N)} dx^\mu \wedge dx^\nu & = & 
\left( 
\begin{array}{cc}
{\rm \fbox{$ F_{\mu\nu}^{(j_1)} dx^\mu \wedge dx^\nu  $}} &                      \\
                     &  {\rm \fbox{$F_{\mu\nu}^{(N-j_1)} dx^\mu \wedge dx^\nu  $}}\\
\end{array}
\right)
 +  F^{({\cal H}_{j_1})}
\end{array} 
\label{FN}
\end{equation}
with
\begin{equation}
F^{({\cal H}_{j_1})} = -\frac{\sin \theta}{2e} d\theta \wedge d\phi \ h_{j_1}
\label{FH}\end{equation}

The $su(\bar N)$ parts and the CSA part of the gauge field
are coupled via the metric functions.
Identifying the non-vanishing functions $\omega_j$ of the $su(N)$
ansatz with the corresponding functions $\bar \omega_i$ of the
non-abelian $su(\bar N)$ ans\"atze, 
\begin{equation}
\gamma_i u_i^2 = \bar \gamma_j \bar u_j^2
\ , \label{scale1} \end{equation}
the asymptotic boundary conditions
for the functions $\bar u_i$,
\begin{equation}
\bar u_i(\infty) = \pm 1
\ , \label{su2l} \end{equation}
yield for the functions 
$u_j= \sqrt{\bar \gamma_i/ \gamma_j} \bar u_i$
the boundary conditions
\begin{equation}
 u_j(\infty) = c_j = \pm \sqrt{\bar \gamma_i/ \gamma_j} 
\ . \label{su2m} \end{equation}

Considering the charge of the solution, we note that as above
the $su(\bar N)$ parts of the solutions are neutral, because the
components $F_{r\theta}^{(\bar{N})}$ and $F_{r\phi}^{(\bar{N})}$
decay like $O(r^{-2})$ asymptotically and the
component  $F_{\theta \phi}^{(\bar{N})}$ decays like $O(r^{-1})$.
In contrast to $F^{(\bar{N})}$,
$F^{({\cal H}_{j_1})}$ does not depend on $r$.
Therefore the charge of the solutions is carried by the CSA part
of the gauge field.
A solution based on
the element $h_{j_1}$ of the CSA then carries
charge of norm $P$ \cite{yasskin},
\begin{equation}  
P^2 = \frac{1}{2}{\rm Tr} \ {h}_{j_1}^2
\ . \label{Pa} \end{equation}
By expanding the element ${h}_{j_1}$ in terms of the
basis $\{\lambda_{n^2-1} \  | \  n=2, \ldots ,N \}$, the charge can also be
directly read off
the expansion coefficients. A particular convenient
expansion involves the charge coefficients
$P_{n^2-1}=\sqrt{\frac{n(n-1)}{2}}$
\begin{equation}
{h}_{j_1} = \sum_{n=2}^N d_{j_1}^n P_{n^2-1} \lambda_{n^2-1} 
\ . \label{Pb} \end{equation}
The  squared norm of the charge,
$P^2$, is a gauge invariant quantity \cite{yasskin},
which enters the equation for the mass function $\mu$.
For the special case $j_1=N-1$,
the ansatz (\ref{amu}) for the gauge field reduces to
the non-abelian spherically symmetric ansatz for the subalgebra $su(N-1)$ 
together with the ansatz for the element of the 
CSA ${h}_{N-1}=P_{N^2-1} \lambda_{N^2-1}$.
Consequently the CSA charge has norm $P=P_{N^2-1}$.
(The case $j_1=1$ is equivalent.)

By applying these considerations again to the subalgebras
$su(\bar N)$ of eq.~(\ref{amu1}), we obtain
the general case for $SU(N)$ EYM theory.
When $M$ gauge field functions are identically zero,
$\omega_{j_m} = 0$, $j_m \in \left\{ j_1, j_2, \ldots , j_M\right\}$, 
$j_1 < j_2 <  \cdots < j_M$,
the ansatz (\ref{amu}) reduces to
\begin{equation}
\begin{array}{ccc}
A_\mu^{(N)} dx^\mu & = & 
\left( 
\begin{array}{cc}
{\rm \fbox{$A_\mu^{(j_1)}dx^\mu$}} &                  \\
                   &\begin{array}{cc}
                    {\rm \fbox{$A_\mu^{(j_2-j_1)}dx^\mu $}}& \\ 
                                           &
                    \begin{array}{cc}
                    {\rm \fbox{$A_\mu^{(j_3-j_2)}dx^\mu $}}& \\ 
                                         &                                
                                            \begin{array}{cc}
                                            \ddots &\\
                   & {\rm \fbox{$A_\mu^{(N-j_M)} dx^\mu $}}
                                            \end{array}                
                     \end{array}
                     \end{array}
\end{array}                     
\right)\\
& &\\
&  &
 + {\cal H}_{j_1, j_2, \ldots, j_M}
\end{array}
\end{equation}
with 
${\cal H}_{j_1, j_2, \ldots, j_M} = \frac{\cos\theta d\phi}{2 e}
{h}_{j_1, j_2, \ldots, j_M}$ and the element 
${h}_{j_1, j_2, \ldots, j_M}$ of the $su(N)$ CSA
\begin{equation}
\begin{array}{c}\!
\left( 
\begin{array}{cc}
\!{\rm \fbox{$(N\!-\!j_1){\bf 1}_{(j_1)}$}}&                  \\
                     &\begin{array}{cc}
                    \!{\rm \fbox{$(N\!-\!j_1\!-\!j_2){\bf 1}_{(j_2-j_1)}$}}& \\ 
                                           &
                    \begin{array}{cc}
                    \!{\rm \fbox{$(N\!-\!j_2\!-\!j_3){\bf 1}_{(j_3-j_2)}$}}& \\ 
                                          &                                
                                            \begin{array}{cc}
                                            \ddots &\\
                     &\!{\rm \fbox{$ \!-\!j_M {\bf 1}_{(N-j_M)}$}}  
                                            \!\end{array}                
                     \!\end{array}
           \!\end{array}                                                        
\!\end{array}  
\!\!\!\!\right)
\\
\!\!\end{array}
\ .  \end{equation}
Here $A_\mu^{(j_1)}, \ldots, A_\mu^{(N-j_M)}$ denote the non-abelian
spherically symmetric ans\"atze for the $su(j_1), \ldots, su(N-j_M)$
subalgebras of $su(N)$,
and ${\cal H}_{j_1,j_2, \ldots, j_M}$ represents the ansatz 
for the element ${h}_{j_1,j_2, \ldots, j_M}$ 
of the $su(N)$ CSA.
The black hole solutions carry CSA charge of norm $P$,
\begin{equation}
P^2 = \frac{1}{2} {\sum_{m=1}^M} (N-j_m)(N-j_{m-1})(j_m-j_{m-1})
\ \ \ (j_0=0)
\ . \label{su2n} \end{equation}
In the special case that
all gauge field functions are identically zero,
an embedded RN solution is obtained with CSA charge of norm $P$ \cite{yasskin},
\begin{equation}
P^2 = \frac{1}{6} (N-1)N(N+1)
\ . \label{pmax} \end{equation}
For the charged $SU(N)$ EYM black holes considered here,
this is the maximal possible norm of the charge.

RN black hole solutions exist only for horizon radius
$x_{\rm H}\ge P$, and
the extremal RN solution has $x_{\rm H}=P$.
As first observed for $SU(3)$ EYM theory \cite{gv},
the same is true for charged non-abelian black holes.
Non-abelian black hole solutions with CSA charge of norm $P$ exist only for 
horizon radii $x_{\rm H} \ge P$.
For extremal black hole solutions ${\cal N}^{'}=0$,
and the coefficient of $u_j^{'}$ in eq.~(\ref{bc5})
vanishes, yielding the boundary conditions
\begin{equation}
\left. (f_{j+1} - f_{j}) u_j \right|_{x_{\rm H}} =0
\ , \label{bc6} \end{equation}
i.e.~$u_j(x_{\rm H})= c_j $,
corresponding to
$\bar u_i(x_{\rm H})= \pm 1 $.
There are no globally regular charged solutions.

\section{Example: Charged SU(5) EYM Black Holes}

We now apply the above general analysis
to $SU(5)$ EYM theory and present magnetically charged $SU(5)$ EYM solutions.
The classification of the charged $SU(5)$
EYM black hole solutions is 
presented for all inequivalent cases (within the ansatz (\ref{amu}))
in Table~1.

Like the neutral non-abelian spherically symmetric solutions
of $SU(N)$ EYM theory, the charged solutions are classified
by the node numbers $n_i$ of their (non-vanishing) gauge field functions
$\bar u_i$.
When the node numbers of one or more gauge field functions
tend to infinity, the solutions approach
limiting solutions with higher norm of the charge.

In Table~2 we show the mass $\mu(\infty)$ of
the lowest solutions of five sequences of black hole solutions,
corresponding to the five inequivalent cases of Table~1,
together with the mass of their limiting solutions.
Shown are case 1a,
case 2a with node structure $(n,0)$,
case 2b with node structure $n$ and $0$,
case 3a with node structure $(n,0,0)$
and case 3b with node structure $(n,0)$ and $0$,
for their extremal horizon $x_{\rm H}=P$ and 
for $x_{\rm H}=\sqrt{20}$.
Their limiting solutions
are the RN solutions with CSA charge of norm $P=\sqrt{20}$ (1a),
$P=\sqrt{19}$ (2a,2b), $P=\sqrt{16}$ (3a)
and $P=\sqrt{18}$ (3b), for $x>P$.
The generalization to all other cases is straightforward.
For instance, for the case 2a the limiting solution 
for a sequence $(i+n,i)$, $i>0$, 
is a non-abelian solution with the same norm of the CSA charge, $P=\sqrt{19}$.

As illustrated in Fig.~1, with increasing $n$ 
the mass $\mu_n(\infty)$ converges exponentially
to the mass $\mu_\infty(\infty)$ of the corresponding limiting solution.
The pattern of convergence of the charged solutions 
is completely analogous to the pattern of convergence of the neutral
solutions as observed previously \cite{kks3},
with the extremal charged solutions playing the role of the 
globally regular neutral solutions.
The extremal solutions fall on straight lines,
whereas the non-extremal black hole solutions 
fall on straight lines only beyond some critical $n$,
for which the horizon radius is getting close to the
innermost node of the gauge field function with $n$ nodes, 
$\bar u_{1,n}$.
The slopes are determined by the particular type of sequence.
Interestingly, we observe
that the slopes of the extremal $SU(2)$ solutions (1a,2b) agree
with the slope of the globally regular $SU(2)$ solutions,
and that their innermost nodes converge with the same
exponent towards a limiting value, slightly smaller than the 
norm of the charge of the limiting solution \cite{lim}.

In Figs.~2a-c we present the lowest odd solutions
of case 3a for the sequence $(n,0,0)$
and extremal horizon $x_{\rm H}=\sqrt{10}$,
also in the interior of the black hole (for $x<x_{\rm H}$).
Figs.~2a-b show the functions $\bar u_1$, $\bar u_2$
and $\bar u_3$, and Fig.~2c shows the charge function
$P(x)$,
$P^2(x)= 2 x \left(\mu(\infty)-\mu(x) \right)
$ \cite{bm,gv}.
The charge functions $P_n(x)$ 
of this sequence of solutions with CSA charge of norm $P=\sqrt{10}$
clearly illustrate that the limiting solution 
carries a CSA charge which has the higher norm $P=4$.

\section{Inclusion of the Dilaton}

Let us end by considering 
$SU(N)$ Einstein-Yang-Mills-dilaton (EYMD) theory.
Analogously to EYM theory, in $SU(2)$ and $SU(3)$ EYMD theory
sequences of neutral globally regular and black hole solutions
exist \cite{eymd,kks3}.
The neutral $SU(2)$ EYMD black hole solutions converge to magnetically charged
Einstein-Maxwell-dilaton (EMD) black hole solutions \cite{emd},
and the neutral $SU(3)$ EYMD solutions converge to limiting
magnetically charged solutions, in which again
one of the two $SU(3)$ gauge field functions is identically zero
\cite{gal,kks3}.
In general, the $SU(N)$ EYM classification remains
valid for $SU(N)$ EYMD theory.
However, the magnetically charged non-abelian black hole solutions
now exist for arbitrary horizon radius $x_{\rm H}>0$,
completely analogous to their abelian (EMD) counterparts.

\newpage

 \newcommand{\rb}[1]{\raisebox{1.5ex}[-1.5ex]{#1}}
\begin{table}[p!]
\begin{center}
\begin{tabular}{|c|cccc|c|l|p{0.75cm}p{0.75cm}p{0.75cm}p{0.75cm}|} \hline
 &  & & & &  & & \multicolumn{4}{ c|}{Cartan subalgebra $^*$} \\
\cline{8-11}
\multicolumn{1}{|c|} { \rb{\#} }&
\multicolumn{4}{ c|} { \rb{$u_j,\ j=1-4$} }&
\multicolumn{1}{ c|} { \rb{$P^2$} }&
\multicolumn{1}{ c|} {\rb{non-abelian  subalgebra}} &
  $\lambda_3$ & $\lambda_8$ & $\lambda_{15}$ & $\lambda_{24}$ \\
 \hline
0  & 0     & 0     & 0     & 0      &
20 &  & 
$P_3$   &  $P_8$    & $P_{15}$  &  $P_{24}$     \\
 \hline
1a & $u_1$ & 0     & 0     & 0     & 
19 & $su(2)$ & 
0   &  $P_8$    & $P_{15}$  &  $P_{24}$   \\
 \hline
2a & $u_1$ & $u_2$ & 0     & 0     & 
16 & $su(3)$ & 
0   &  0    & $P_{15}$  &  $P_{24}$  \\
2b & $u_1$ & 0     & $u_3$ & 0     & 
18 & $su(2)\oplus su(2)$ &
0   & $\frac{4}{3}P_8$    & $\frac{2}{3}P_{15}$  &  $P_{24}$ \\
 \hline
3a & $u_1$ & $u_2$ & $u_3$ & 0     & 
10 & $su(4)$  &
0   & 0    & 0  &  $P_{24}$ \\
3b & $u_1$ & $u_2$ & 0     & $u_4$ & 
15 & $su(3)\oplus su(2)$ &
0   & 0    & $\frac{5}{4}P_{15}$  &  $\frac{3}{4}P_{24}$\\
\hline
\end{tabular}
\end{center} 
\vspace{1.cm} 
{\bf Table 1}\\
Classification of the magnetically charged black hole solutions of $SU(5)$
EYM theory.
Shown are the non-vanishing gauge field functions (denoted by $u_j$)
and the identically vanishing gauge field functions (denoted by zero),
the square of the norm of the charge of the black hole solutions, $P^2$,
and the subalgebras of the solutions including the non-abelian neutral
part and the CSA charged part, with the
coefficients (in the given basis {*}) of the corresponding 
element of the CSA.
\end{table}

\clearpage

\newpage
\begin{table}[p!]
\begin{center}
\begin{tabular}{|c|ccccc|} \hline
\multicolumn{1}{|c|} { $ $ }&
\multicolumn{5}{ c|} {  $\mu(\infty)$ }\\
$ $ & 1a     & 2a     & 2b     & 3a     & 3b     \\
\hline
$n/x_{\rm H}$& $\sqrt{19}$ & $\sqrt{16}$ & $\sqrt{18}$ &
 $\sqrt{10}$ & $\sqrt{15}$ \\
 \hline
$1$      & 4.44954 & 4.25227 & 4.33571 &  3.69884 & 4.13289 \\
$2$      & 4.46834 & 4.33108 & 4.35500 &  3.90097 & 4.21401 \\
$3$      & 4.47152 & 4.35195 & 4.35826 &  3.96889 & 4.23549 \\
$4$      & 4.47204 & 4.35719 & 4.35880 &  3.99041 & 4.24088 \\
$5$      & 4.47212 & 4.35848 & 4.35888 &  3.99706 & 4.24221 \\
$\infty$ & 4.47214 & 4.35890 & 4.35890 &  4.0     & 4.24264 \\
 \hline
\multicolumn{1}{|c|} { $ $ }&
\multicolumn{5}{ c|} {  $x_{\rm H}=\sqrt{20}$ } \\
 \hline
$1$      & 4.45085 & 4.27291 & 4.34103 & 3.83752 & 4.16591 \\
$2$      & 4.46919 & 4.34438 & 4.35861 & 3.98865 & 4.23561 \\
$3$      & 4.47182 & 4.35841 & 4.36027 & 4.02026 & 4.24735 \\
$4$      & 4.47211 & 4.36018 & 4.36033 & 4.02446 & 4.24845 \\
$5$      & 4.47213 & 4.36032 & 4.36033 & 4.02488 & 4.24852 \\
$\infty$ & 4.47214 & 4.36033 & 4.36033 & 4.02492 & 4.24853 \\
 \hline
\end{tabular}
\end{center} 
\vspace{1.cm} 
{\bf Table 2}\\
The dimensionless mass $\mu(\infty)$
of the lowest black hole solutions
of the sequences corresponding to
case 1a of Table~1,
case 2a with node structure $(n,0)$,
case 2b with node structure $n$, $0$,
case 3a with node structure $(n,0,0)$
and case 3b with node structure $(n,0)$, $0$
for their extremal horizons
and for $x_{\rm H}=\sqrt{20}$.
For each sequence the corresponding limiting value
is shown in the last row (denoted by $\infty$).

\end{table}

\clearpage

\begin{figure}
\centering
\epsfysize=11cm
\mbox{\epsffile{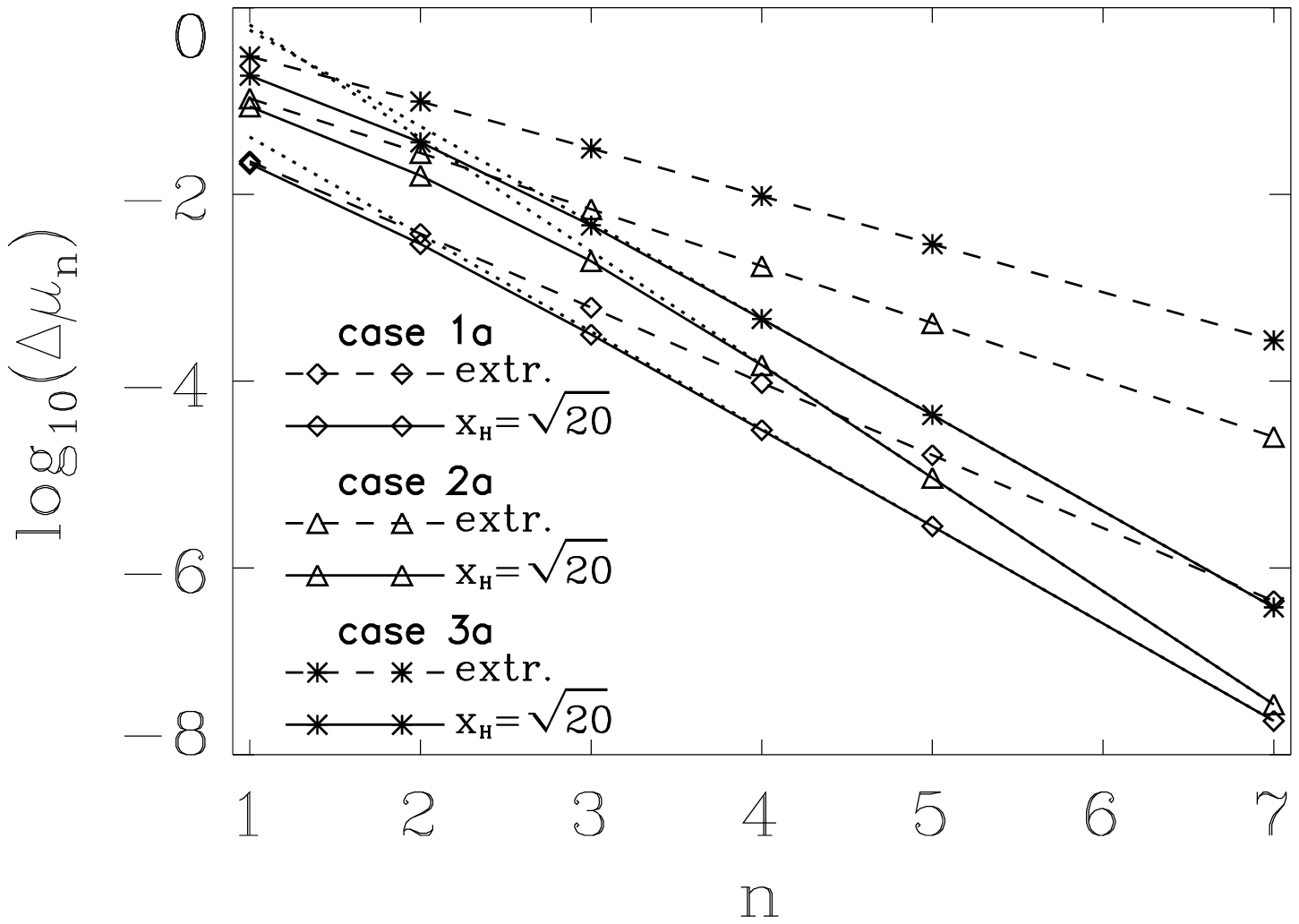}}
\end{figure}
\noindent
{\bf Figure~1:\\ }
The logarithm of the absolute deviation from the limiting solution 
$\Delta \mu_n = \mu_\infty (\infty)-\mu_n (\infty)$ for the masses of 
the cases 1a, 2a and 3a of Table~1
as a function of the node number $n$ for
the extremal solutions (dashed lines) and the non-extremal solutions 
(solid lines) with horizon at $x_{\rm H}=\sqrt{20}$. 
The dotted lines indicate the asymptotic behaviour of 
the non-extremal solutions.
\newpage
\begin{figure}
\centering
\epsfysize=11cm
\mbox{\epsffile{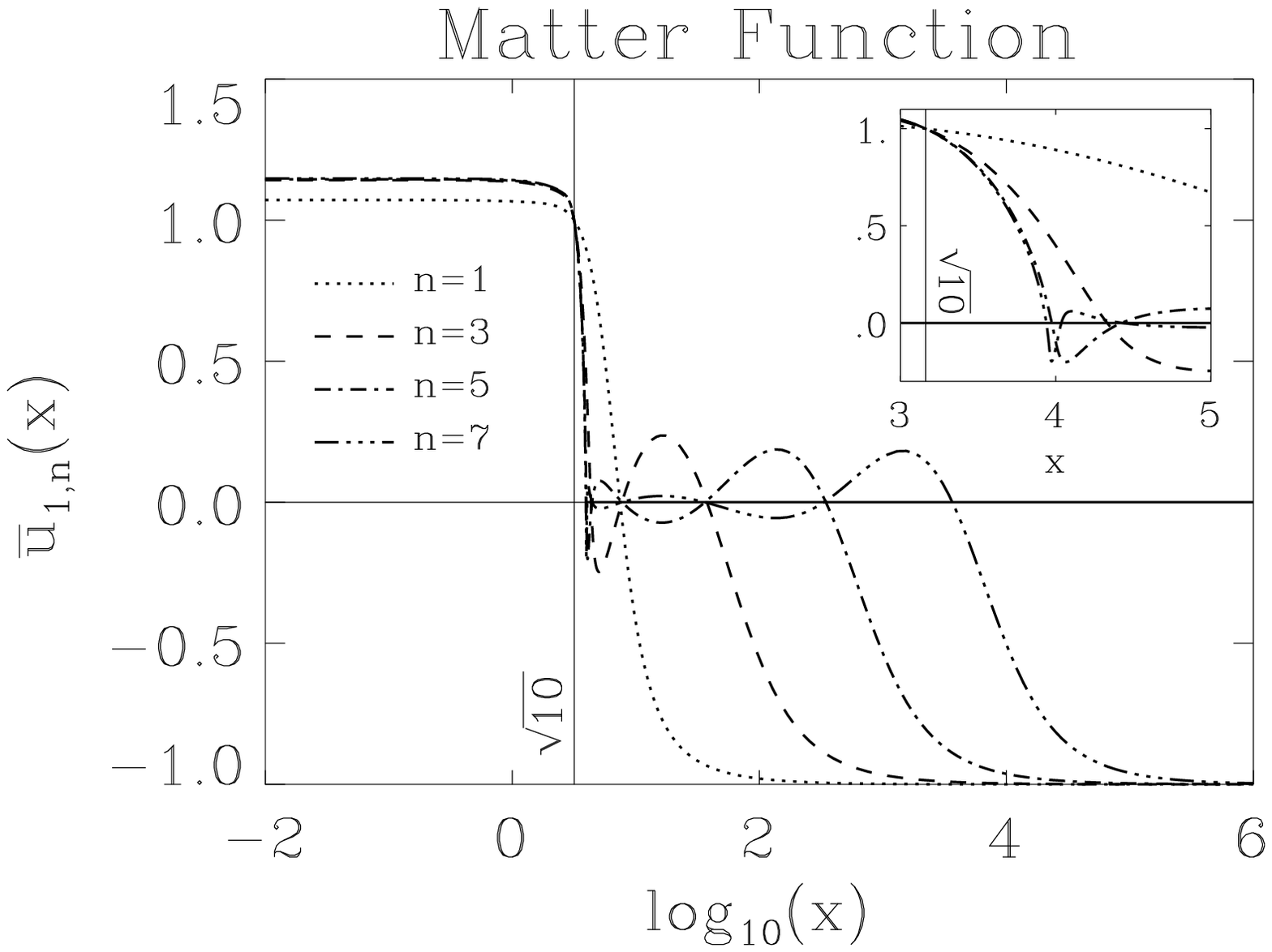}}
\end{figure}
\noindent
{\bf Figure~2a:\\ }
The matter functions $\bar u_{1,n}(x)$ for case 3a of Table~1
with the node structure $(n,0,0)$
and extremal event horizon $x_{\rm H} = \sqrt{10}$.
\newpage
\begin{figure}
\centering
\epsfysize=11cm
\mbox{\epsffile{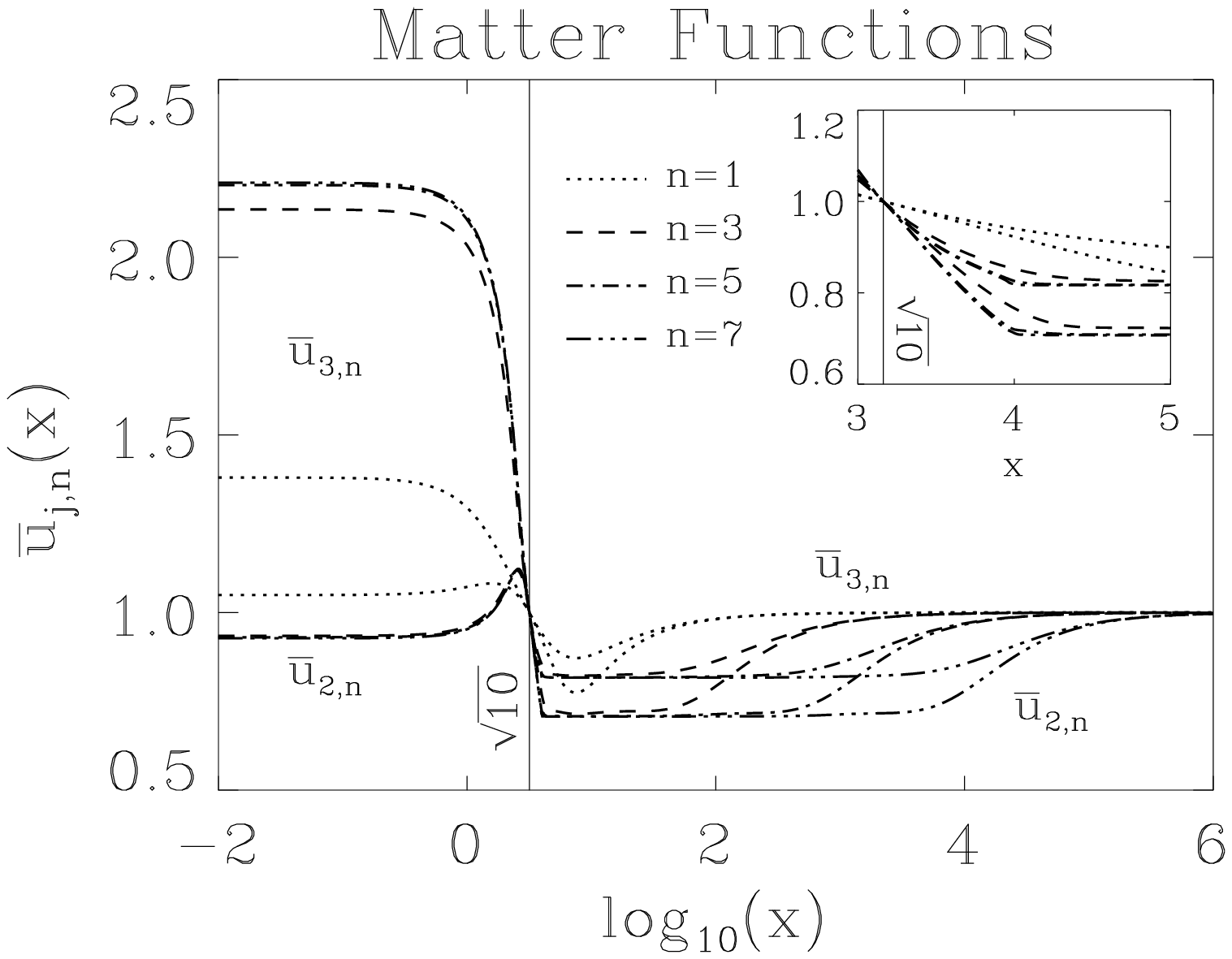}}
\end{figure}
\noindent
{\bf Figure~2b:\\ }
The same as Fig.~2a for the matter functions 
$\bar u_{2,n}(x)$ and $\bar u_{3,n}(x)$.
\newpage
\begin{figure}
\centering
\epsfysize=11cm
\mbox{\epsffile{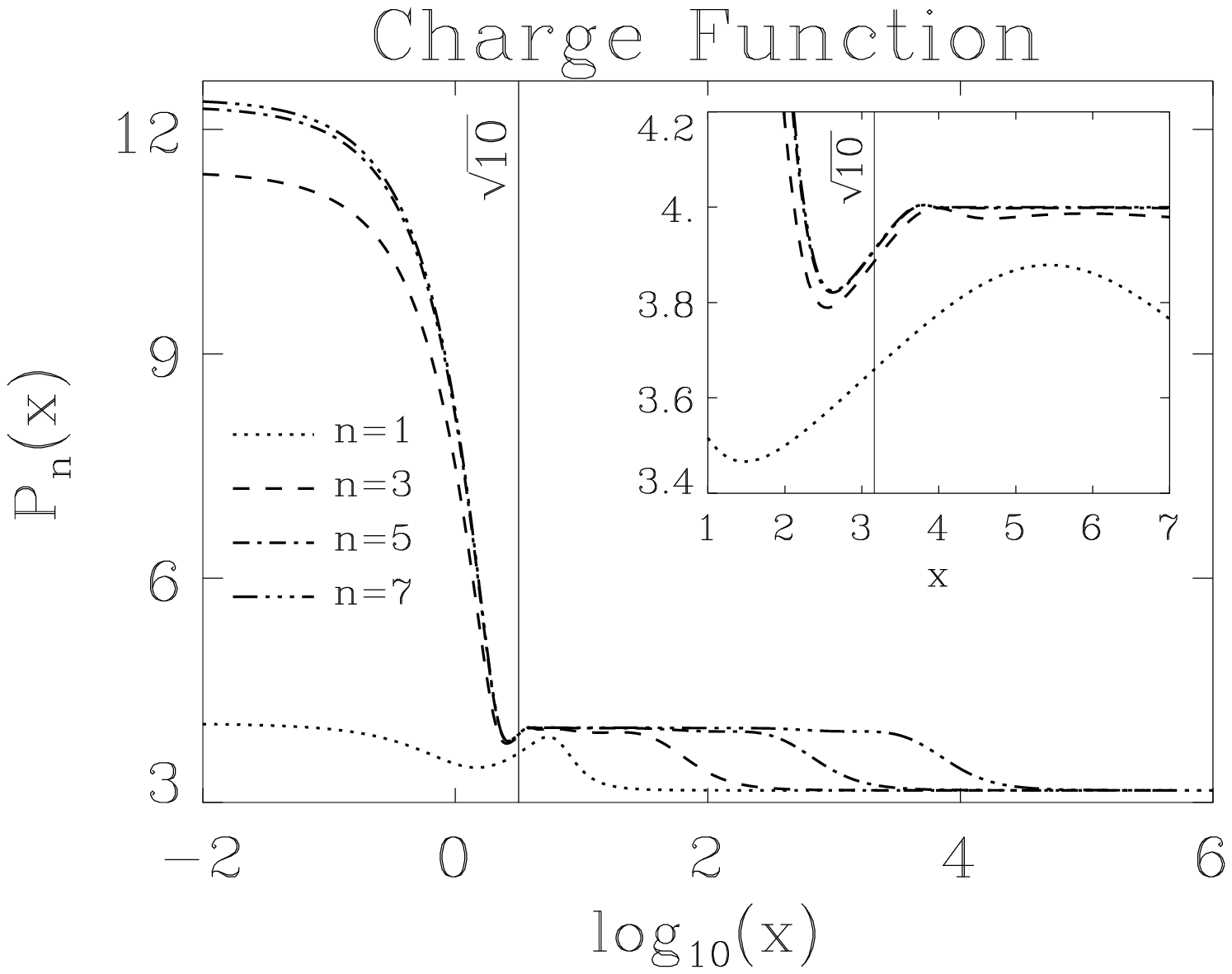}}
\end{figure}
\noindent
{\bf Figure~2c:\\ }
The same as Fig.~2a for the charge function $P_n(x)$.
\newpage
\end{document}